\documentclass[aps,prd,10pt,notitlepage,nofootinbib,superscriptaddress,showkeys,showpacs]{revtex4-1}
\usepackage{amsmath,amssymb,amsthm,latexsym}
\usepackage[english]{babel}
\usepackage{graphicx,color}
\usepackage{xspace}
\usepackage{graphicx}
\usepackage{pifont,dsfont}
\usepackage{marvosym}
\usepackage{slashed}
\usepackage{subfigure}

\newcommand{\be}{\begin{equation}}
\newcommand{\ee}{\end{equation}}
\newcommand{\bqa}{\begin{eqnarray}}
\newcommand{\eqa}{\end{eqnarray}}
\newcommand{\bea}{\begin{eqnarray}}
\newcommand{\eea}{\end{eqnarray}}

\newcommand{\N}{\mathds{N}}

\DeclareMathOperator{\tr}{tr}

\begin{document}

\title{\Large \bf Coupling of hard dimers to dynamical lattices via random tensors}

\author{{\bf Valentin Bonzom}}\email{vbonzom@perimeterinstitute.ca}
\author{{\bf Harold Erbin}}\email{harold.erbin@gmail.com}
\affiliation{Perimeter Institute for Theoretical Physics, 31 Caroline St. N, ON N2L 2Y5, Waterloo, Canada}

\date{\small\today}

\begin{abstract}
We study hard dimers on dynamical lattices in arbitrary dimensions using a random tensor model. The set of lattices corresponds to triangulations of the d-sphere and is selected by the large N limit. For small enough dimer activities, the critical behavior of the continuum limit is the one of pure random lattices. We find a negative critical activity where the universality class is changed as dimers become critical, in a very similar way hard dimers exhibit a Yang-Lee singularity on planar dynamical graphs. Critical exponents are calculated exactly. An alternative description as a system of ``color-sensitive hard-core dimers'' on random branched polymers is provided.
\end{abstract}

\medskip

\noindent  Pacs numbers: 02.10.Ox, 04.60.Gw, 05.40-a

\maketitle

\section{Introduction}

Statistical mechanics on random, or dynamical, lattices consists in averaging a lattice model over some ensemble of graphs. One usually works in a grand-canonical ensemble, where the lattice chemical potential is the weight for adding a lattice site. The thermodynamic limit is the regime where the partition function is dominated by graphs with a huge number of sites, i.e. when the expectation value of the number of sites goes to infinity. In this regime, the partition function develops some singular behavior and one says that the lattice becomes critical \cite{Kazakov:1985ds, mm}.

Hard dimers are objects which can be attached to links of a lattice with some exclusion rule. Together with Ising spins \cite{Kazakov:1986hu}, they are one of the first systems of statistical mechanics which have been studied on random two-dimensional lattices \cite{yang-lee-staudacher}. Using a matrix model formulation, the partition function can be calculated exactly. In the continuum limit, dimers and lattices can become critical together, and this gives rise to the Yang-Lee singularity, a non-unitary conformal field theory (CFT) with central charge $c=-22/5$ coupled to two-dimensional quantum gravity \cite{ginsparg-moore-ising, gross-migdal-ising}. This universality class is generically observed in models of hard objects \cite{hard-objects}. Thus, hard dimers provide a typical model for multicritical behaviors on random 2d surfaces, which has paved the way to higher order multicritical behaviors \cite{kazakov-matter, brezin-ising, graph-combinatorics-difrancesco, mm-review-difrancesco}.

Random rank-$d$ tensor models have been introduced to generate random lattices in dimensions $d>2$ \cite{ambjorn-3d-tensors, sasakura-tensors, gross-tensors}. A class of such models which relies on the use of an ensemble of colored graphs \cite{Gur3, GurRiv, Gur4}, is now known to have a continuum limit in the large $N$ scaling limit, $N$ being the size of each tensor entry, \cite{Bonzom:2011zz}. Beyond the criticality of pure random lattices, other universality classes have been found \cite{toy-double-scaling, uncoloring}, but little has been said about their physical origin. We also refer to \cite{EDT} for the appearance of non-universal behaviors, \cite{coloredreview} for a detailed review of tensor models and \cite{uncoloring} for a synthetic presentation of the status of the field based on the result of \cite{universality-gurau}.

To understand the physics of these multicritical behaviors, it was proved fruitful in two dimensions to consider the Ising model and hard dimers. Using a two-matrix model, it was found that Ising spins in the continuum have the universal behavior of the unitary Ising CFT (central charge $c=1/2$) coupled to quantum gravity \cite{Kazakov:1986hu, Boulatov:1986sb}. However in higher dimensions, using the large $N$ limit of a colored tensor model, we found that there is no phase transition at finite temperature \cite{ising-colored} in agreement with previous numerical simulations \cite{ambjorn-3d-ising}.

The study of dimers coupled to random tensors has been initiated in \cite{Bonzom:2012sz}. There the model features dimers with a modified exclusion principle (they are not completely hard-core) and exhibits in a remarkably simple fashion multicritical behaviors. In particular an example of a phase transition was proposed at $d=6$. Dimers have also been recently considered to reach multicritical points in two-dimensional causal dynamical triangulations \cite{ambjorn-2dCDT-dimers} (see also \cite{atkin-multicriticalCDT}).

In the present paper we continue this investigation with truly hard-core dimers. In Sec. \ref{sec:dimers-spheres} we review the two-dimensional case and present the generalization to arbitrary dimensions in terms of tensors. The model is solved to leading order in the large $N$ limit in Sec. \ref{sec:critical-dimers}. While the analysis holds for any $d$ we have chosen to give systematic details on the case $d=3$. Our result is similar to the two-dimensional situation: dimers become critical for some negative activity, sending the continuum limit to a universality class different from pure random lattices. This universality class is the one already observed in \cite{Bonzom:2012sz, uncoloring}. Its entropy exponent $\gamma$, which characterizes the proliferation of microstates (graphs dressed with dimers) in the continuum limit, is the case $m=3$ of the series $\gamma=1-1/m$ of \cite{Bonzom:2012sz, uncoloring}.

The dominant graphs of our model are known to be in one-to-one correspondence with branched polymers \cite{Bonzom:2011zz}. This provides an alternative representation of the model where hard-core dimers are mapped to \emph{color-sensitive hard-core dimers}, as we explain in Sec. \ref{sec:mapmelons-trees}.

In Sec. \ref{sec:moredimers} we revisit the model of \cite{Bonzom:2012sz} through a different generating function which is more natural from the view of the thermodynamic limit. The qualitative conclusions are the same and so are the exact entropy exponents $\gamma=1-1/m$, which are also the same as for multicritical branched polymers \cite{BP-ambjorn}. We discuss the case $d=3$ in details.


\section{Hard dimers on random spherical lattices} \label{sec:dimers-spheres}

A dimer on a lattice $G$ is a bond on a link between two nodes. It has an activity $z$ which is the weight for putting a dimer on a link. In addition, one often considers some exclusion rules, i.e. constraints preventing the dimerization of all lattice links. These exclusion rules create some frustration and there are usually many equivalent ways to put a fixed number of dimers on the lattice. A dimer configuration is a set of lattice links carrying dimers, not necessarily a covering. Denote $\mathcal D(G)$ the set of configurations on $G$ allowed by the exclusion rules. Then the partition function is
\be
\Xi_{G}(z) = \sum_{D\in\mathcal D(G)} z^{|D|},
\ee
where $|D|$ is the number of dimers in the configuration $D$. Making the lattice dynamical means that we now sum over a set of graphs, with a fixed number of vertices,
\be
\Xi_n(z) = \sum_{\{G_n\}} \Xi_{G_n}(z).
\ee
Introducing some chemical potential $g$ for the number of lattice sites we get a grand-canonical ensemble
\be
\Xi(z,g) = \sum_{n=0}^\infty \Xi_n(z)\,g^n.
\ee
Assuming this generating function has a finite radius of convergence $g_c(z)$, depending on the activity $z$, the thermodynamic limit is the regime where analyticity is lost, so that the grand-canonical partition function is dominated by graphs with an infinite number of sites. Close to $g_c(z)$, the expectation value for the number of sites calculated in the grand canonical ensemble goes like $\langle n \rangle \sim 1/|g-g_c(z)|$.

The large $n$ asymptotics of $\Xi_n(z)$ is expected to be of the form $\Xi_n(z) \sim A\,n^{\gamma-3}\,[g_c(z)]^{-n}$ where $\gamma$ is the \emph{entropy exponent}. In the grand-canonical ensemble, it translates to the following singular behavior $\Xi(z,g) \sim (g-g_c(z))^{2-\gamma}$.

\subsection{The 2d case: Hard dimers on random planar graphs} \label{sec:hard-dimers-2d}

Such generating functions on two-dimensional graphs can be computed as the free energy of some hermitian matrix models. Taking the graphs in $\{G_n\}$ to be 4-valent graphs on the 2-sphere, hence planar, we have \cite{yang-lee-staudacher}
\be
e^{N^2 \Xi(z,g)} =\int dA\,dB\ \exp\left[ -N\,\tr\ \frac12 A^2 - \frac{g}4 A^4 +\frac12 B^2 -g\sqrt{z}\,B\,A^3\right],
\ee
at the leading order for large $N$. Using the saddle point equation for the spectral density, one finds that for most values of the dimer activity $z$ the continuum limit lies in the universality class of pure random lattices (i.e. pure 2d quantum gravity). The singular behavior of $\Xi(z,g)$ reads
\be
\Xi_{\text{sing}}(z,g) \sim (g-g_c(z))^{2-\gamma},\quad \text{with $\gamma=-1/2$}.
\ee
However, when the activity reaches a critical value $z_c=-1/10$, dimers become critical too and change the universality class,
\be
\Xi_{\text{sing}}(z_c,g) \sim (g-g_c(z_c))^{2-\gamma},\quad \text{with $\gamma=-1/3$}.
\ee
When one approaches the singularity along the curve of the continuum limit $g_c(z)$, another critical exponent is found
\be
\frac{d}{dz} \log g_c(z) \sim (z-z_c)^{1/2}.
\ee

This critical point is identified with the Lee-Yang CFT, of central charge $c=-22/5$, coupled to Liouville gravity \cite{hard-objects, graph-combinatorics-difrancesco, mm-review-difrancesco}. A universality check can be done, using planar triangulations instead of quadrangulations, leading to the same critical exponents.

\subsection{Dynamical spherical triangulations in arbitrary dimensions via random tensors} \label{sec:pure-tensors}

We now propose to generalize this approach to lattices in higher dimensions, using random tensors instead of random matrices. The most naive generalization of matrices to tensors, introduced in 1991, is not known to admit a large $N$ expansion. Instead we need a slightly refined framework which does support a large $N$ expansion dominated by some triangulations of the $d$-sphere, known as colored tensor models.

We introduce a collection of $(d+1)$ tensors $(T^{(i)}_{a_1\dotsb a_d})_{i=0,\dotsc,d}$, where $(i)$ is called the color index and the indices $a_1,\dotsc,a_d$ are the usual tensor indices, ranging from $1$ to $N$. These tensors are complex and their conjugate are denoted $\bar T^{(i)}$. These tensors are useful to generate \emph{bipartite $(d+1)$-colored graphs}. Lines of a graph are generated by propagators between some $T^{(i)}$ and $\bar T^{(i)}$ and hence carry a color index. This leads to the quadratic part of the tensor action
\be
S_{\text{quad}} = \sum_{i=0}^d \sum_{a_1,\dotsc,a_d} T^{(i)}_{a_1\dotsb a_d}\ \bar T^{(i)}_{a_1\dotsb a_d}.
\ee
The sites of each graph are chosen of degree $(d+1)$ with distinct adjacent colors. A set of `black' vertices is generated by
\be
S_+ = \sqrt{g} \sum_{\{n_{ij}\}} \prod_{i=0}^d T^{(i)}_{n_{i i-1}\dotsb n_{i0} n_{id}\dotsb n_{ii+1}},
\ee
and a set of `white' vertices by
\be
S_- = \sqrt{g} \sum_{\{n_{ij}\}} \prod_{i=0}^d \bar T^{(i)}_{n_{i i-1}\dotsb n_{i0} n_{id}\dotsb n_{ii+1}}.
\ee
Colored graphs obviously possess vertices and lines but also a richer structure due to the coloring. One can for example define faces as the maximally connected subgraphs with exactly two colors. These faces are interesting because, just like in random matrix models, the amplitude of a graph has a free sum per face, hence part of its scaling going like $N^F$, where $F$ is the number of faces.

To have a well-defined large $N$ limit, one must compensate in some way the above scaling with the number of faces. To this aim, it is necessary to re-scale the action with some power of $N$, which helps peak the integral appropriately, as follows
\be
e^{-N^d F(g)} = \int [dT^{(i)}\,d\bar T^{(i)}]\ \exp -N^{d/2} \Bigl( S_{\text{quad}}+S_++S_-\Bigr)
\ee
Then the amplitude of a connected graph scales like $N^{F-\frac{d(d-1)}{4} V}$, where $V$ is the number of vertices. It can be shown that $F-\frac{d(d-1)}{4} V$ is bounded by $d$ and writes $F-\frac{d(d-1)}{4} V = d-\frac{2}{(d-1)!}\omega(G)$, where $\omega(G)$ is a positive integer called the \emph{degree} of the graph $G$. This provides the free energy of the model $F(g)$ with a large $N$ expansion
\be
F(g) = \sum_{\omega \in \frac{(d-1)!}{2}\N} N^{-\frac{2}{(d-1)!}\omega}\ f_\omega(g) = f_0(g) + \frac{1}{N}\,f_{\omega = \frac{(d-1)!}{2}}(g) + \dotsb.
\ee
Each $f_\omega(g)$ is the free energy at fixed degree $\omega$. For our purpose it is not necessary to give the definition of the degree of a colored graph, as we will only focus on graphs of vanishing degree and provide a precise characterization of them. Note however that this framework makes sense for any $d\geq 2$. When $d=2$, i.e. tensors reduce to matrices, the degree $\omega$ is exactly the \emph{genus} of the graph, as expected.

As in 2d, the colored graphs generated by such colored tensor models have a natural topological interpretation in dimension $d$. To each vertex of degree $(d+1)$, one associates a simplex of dimension $d$. It has $(d+1)$ simplices on its boundary, which are represented on the graph by the lines meeting at the vertex. Hence, these boundary simplices are identified by a color $(i)$. A $(d-2)$-dimensional sub-simplex of a $d$-simplex is shared by exactly two $(d-1)$-simplices on the boundary, say of colors $(i)$ and $(j)$, and is therefore uniquely identified by the pair of colors $(ij)$. This identification scheme of sub-simplices in a simplex works for any dimension and is made possible thanks to colors. When a line of the graph joins two vertices, this means that two $d$-simplices are glued by identifying one of their boundary $(d-1)$-simplex. This simplex is again uniquely identified by the color of the line. Moreover, colors provide an unambiguous way to glue the two simplices, using the unique gluing which respects all the induced colorings on the sub-simplices. This prescription associates a topology to any colored graph and it can be showed that all the graphs generated here are topological pseudo-manifolds in dimension $d$.

\begin{figure}
 \includegraphics[scale=0.4]{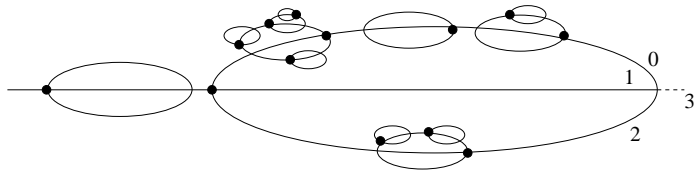}
 \caption{\label{fig:typical-melon}A typical melonic graph for $d=3$ with two external legs. Notice that vertices come by pair (a black and a white, connected by $d$ lines with melonic insertions). }
\end{figure}

Graphs of degree zero have the topology of the $d$-sphere. Moreover, their combinatorics can be described quite precisely \cite{Bonzom:2011zz}. They are found as the graphs which maximize the number of faces at fixed number of vertices. For $d> 2$, there is a way to locally maximize the number of faces by inserting on any line a patch with two vertices connected by $d$ lines. This contributes to $d(d-1)/2$ faces with exactly two vertices and does not change the degree of the graph. The family of closed \emph{melonic graphs} is the family of graphs obtained by recursive insertions of this patch on any line starting from the closed graph with two vertices connected by $(d+1)$ lines. A typical melonic graph is depicted in Fig. \ref{fig:typical-melon}.

\begin{figure}
 \includegraphics[scale=0.6]{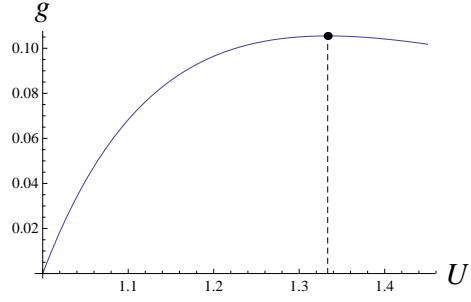}
 \caption{\label{fig:gdeU-purelattice} The equation $U=1+g U^{d+1}$ in the plane $(g,U)$ for $d=3$. The Gaussian model is for $g=0, U=1$. The continuum limit is the black dot where $\partial g/\partial U=0$ so that $(U-U_c)\sim \sqrt{g_c-g}$. One has $U_c = (d+1)/d$ and $g_c = d^d/(d+1)^{d+1}$.}
\end{figure}

Equivalently, the generating function for melonic graphs satisfies some algebraic equation. Denote $U$ the full connected 2-point function, $\langle T^{(i)}_{a_1\dotsb a_d} \bar T^{(j)}_{b_1 \dotsb b_d}\rangle = N^{-d/2} U \delta^{ij} \prod_c \delta_{a_c b_c}$, and similarly $\Sigma$ the 1-particle-irreducible 2-point function. One has as usual $U=1/(1-\Sigma)$. In the melonic sector, one gets an additional equation to close the system on $U$,
\be \label{melonic1PI}
\Sigma = g\ \begin{array}{c}\includegraphics{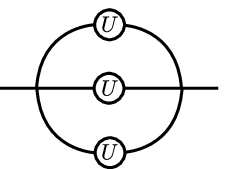}\end{array} =
g\ U^d,
\ee
This means that all contributions to $\Sigma$ factorize on $d$ internal lines connecting the left and right vertices, with insertions of melonic 2-point graphs. One gets all contributions by inserting the full 2-point function itself $U$ on each line. 
As a result,
\be
U = 1+g\,U^{d+1}.
\ee
The solution to this equation is represented in the plane $(g,U)$ in Fig. \ref{fig:gdeU-purelattice}. Starting from the Gaussian model for $g=0, U=1$, one follows the curve of the solution until a stationary point is reached. There $g-g_c$ goes quadratically with $(U-U_c)$ as the first derivative vanishes. Therefore $(U-U_c) \sim \sqrt{g-g_c}$. This singular behavior defines the thermodynamic limit and one extracts the entropy exponent $\gamma_{\text{melons}}=1/2$.

\subsection{Hard dimers on melonic triangulations via a tensor model} \label{sec:dimers-melons}

The above model can be extended to include dimer configurations dressing the dynamical lattices. A set of tensors $(\chi^{(i)}_{a_1\dotsb a_d})_{i=0,\dotsc,d}$, together with their complex conjugate, is introduced to represent a dimer on a link of color $(i)$. The hard-core repulsion is then fully encoded into the interaction of these tensors with the ones for pure lattices. As only a single dimer can touch a site of the lattice, we need the following interaction
\be
S^{\text{dimer}(i)}_{+} = \sqrt{gz} \sum_{\{n_{ij}\}} \chi^{(i)}_{n_{i i-1}\dotsb n_{i0} n_{id}\dotsb n_{ii+1}}
\prod_{j\neq i} T^{(j)}_{n_{j j-1}\dotsb n_{j0} n_{jd}\dotsb n_{jj+1}},
\ee
and similarly with the complex conjugate $S^{\text{dimer}(i)}_{-}$. In words, one removes the tensor $T^{(i)}$ of color $(i)$ in the interaction $S_\pm$ and put the dimer tensor $\chi^{(i)}$ of the same color instead, and weights this vertex with $\sqrt{gz}$. Dimers also need a quadratic part $S_{\text{quad}}^{\text{dimer}}$ which is the same as $S_{\text{quad}}$ with $\chi$ instead of $T$.

The partition function is
\be
Z = e^{-N^d \Xi(z,g)} = \int [dT^{(i)} d\chi^{(i)}\,d\bar T^{(i)} d\bar\chi^{(i)}]\ \exp -N^{d/2} \Bigl( S_{\text{quad}} + S_{\text{quad}}^{\text{dimer}} + S_+ + S_- + S^{\text{dimer}}_{+}+S^{\text{dimer}}_{-}\Bigr),
\ee
where $S^{\text{dimer}}_{\pm} = \sum_{i=0}^d S^{\text{dimer}(i)}_{\pm}$. We have denoted the free energy of the model $\Xi(z,g)$ because it is indeed the grand-canonical partition function for hard dimers on the set of random bipartite $(d+1)$-colored graphs, with lattice chemical potential $\sqrt{g}$ and dimer activity $z$. This grand-canonical partition function has a degree expansion which is completely independent of the dimer configurations and hence exactly the same as for pure random lattices.

In particular, at degree zero,
\be
\Xi_0(z,g) = \sum_{\substack{\text{melonic graphs $G_{V}$}\\ \text{dressed with $D\in \mathcal{D}(G_{V})$}}} g^{V/2}\ z^{|D|}.
\ee

\section{Critical behaviors} \label{sec:critical-dimers}

We now somehow repeat the procedure of Sec. \ref{sec:pure-tensors}, with details, to solve the model at the melonic order. Due to coloring, it is guaranteed that the full connected 2-point functions can be written
\begin{gather}
\langle T^{(i)}_{a_1\dotsb a_d}\,\bar T^{(i)}_{a_1\dotsb a_d} \rangle = N^{-d/2}\,U,\qquad \langle \chi^{(i)}_{a_1\dotsb a_d}\,\bar \chi^{(i)}_{a_1\dotsb a_d} \rangle = N^{-d/2}\,W,\\
\langle \chi^{(i)}_{a_1\dotsb a_d}\,\bar T^{(i)}_{a_1\dotsb a_d} \rangle = \langle T^{(i)}_{a_1\dotsb a_d}\,\bar \chi^{(i)}_{a_1\dotsb a_d} \rangle = N^{-d/2}\,V,
\end{gather}
all other components vanishing.

First we use some Schwinger-Dyson equations to relate the free energy to these 2-point functions.
\be
\frac{1}{Z\ N^d} \int [dT^{(i)} d\chi^{(i)}\,d\bar T^{(i)} d\bar\chi^{(i)}] \sum_{a_1,\dotsc,a_d} \frac{\partial}{\partial T^{(i)}_{a_1\dotsb a_d}}\Bigl( T^{(i)}_{a_1\dotsb a_d}\ e^{-N^{d/2}(S_{\text{quad}}(T,\bar T) + S_{\text{quad}}(\chi,\bar \chi) + S_+ + S_- + S^{\text{dimer}}_+ +S^{\text{dimer}}_-)} \Bigr) = 0.
\ee
Since each term of the action is at most linear in each $T^{(i)}$, evaluating the derivatives yields
\be
1 - U - N^{-d/2}\,\langle S_+ + S^{\text{dimer}}_+ - S^{\text{dimer}(i)}_{+}\rangle = 0.
\ee
We sum this equation with the same one coming from $\bar T^{(i)}$ and also sum over colors,
\be
1 - U - \frac{N^{-d/2}}{2}\,\langle S_+ + S_- + S^{\text{dimer}}_+ + S^{\text{dimer}}_-\rangle + \frac{N^{-d/2}}{2(d+1)} \langle S^{\text{dimer}}_+ + S^{\text{dimer}}_-\rangle = 0.
\ee
Then the connected expectation values in the above equation are recast as derivatives of the free energy to give
\be
1 - U - g\,\frac{\partial\, \Xi}{\partial g} + \frac{z}{d+1}\,\frac{\partial\, \Xi}{\partial z} = 0.
\ee
This reasoning is repeated from a Schwinger-Dyson equation with a $\chi^{(i)}$ insertion instead to get
\be
1 - V - \frac{z}{d+1}\,\frac{\partial \, \Xi}{\partial z} =0.
\ee
By combining the two above equations, we are able to write the derivatives of the free energy as
\be \label{SD}
g\,\frac{\partial\,\Xi}{\partial g} = 2 - U - V,\qquad z\,\frac{\partial \, \Xi}{\partial z} = (d+1)(1-V),
\ee
which reduce the problem to finding $U$ and $V$ in the melonic sector.

\subsection{The thermodynamic limit}

We denote $\Sigma = \left(\begin{smallmatrix} \Sigma_U &\Sigma_W\\ \Sigma_W &\Sigma_V\end{smallmatrix}\right)$ the matrix of the 1PI 2-point functions. Its geometric series is exactly the matrix of the full connected 2-point functions,
\be
\begin{pmatrix} U &W\\W &V\end{pmatrix} \begin{pmatrix} 1- \Sigma_U & -\Sigma_W \\ -\Sigma_W &1-\Sigma_V \end{pmatrix} = \begin{pmatrix} 1 &0\\ 0 &1\end{pmatrix}.
\ee
Since they are symmetric matrices, one can just consider three real equations like
\begin{align}
U(1-\Sigma_U) - W\,\Sigma_W &= 1,\label{getU}\\
V(1-\Sigma_V) - W\,\Sigma_W &= 1, \label{getV}\\
W(1-\Sigma_V) - U\,\Sigma_W&= 0. \label{getW}
\end{align}
Therefore we just have to find some extra equations in the melonic sector which enable to close the system on $U, V, W$. The melonic sector is characterized by the fact that all contributions comes from 2-point insertions which are themselves melonic on any line. To generalize \eqref{melonic1PI} in the presence of dimers, we just have to list all the possible ways of putting dimers in the graph of \eqref{melonic1PI}, weighted appropriately with the dimer activity and some combinatorial factors counting the equivalent ways to put a dimer. This gives for $\Sigma_U$
\begin{align}
&\begin{aligned}\Sigma_U =
g\ \begin{array}{c} \includegraphics{selfE_U_1.eps} \end{array} + dg\sqrt{z}\ &\left[ \begin{array}{c}\includegraphics{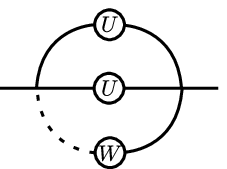}\end{array} + \begin{array}{c}\includegraphics{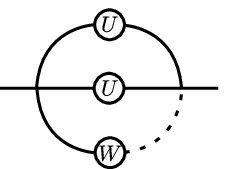}\end{array}\right]\\ &+ dgz\ \begin{array}{c}\includegraphics{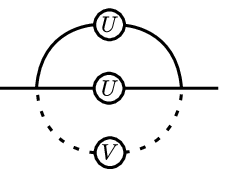}\end{array} + d(d-1)gz \begin{array}{c}\includegraphics{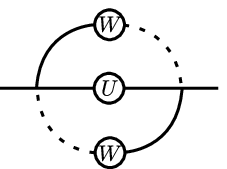}\end{array},\end{aligned}\\
&\phantom{\Sigma_U} = g U^d + 2dg \sqrt{z}\,W\,U^{d-1} + dgz\,V\,U^{d-1} + d(d-1)gz\,W^2\,U^{d-2},\label{sigmaU}
\end{align}
where we have represented graphically the contributions at $d=3$ with dimers as dashed lines. For $\Sigma_V$, one cannot make any dimer 2-point insertion so that there is only one kind of contributions,
\be
\Sigma_V = gz\ \begin{array}{c}\includegraphics{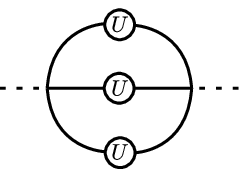}\end{array}
= gz\,U^d,\label{sigmaV}
\ee
And finally
\be
\Sigma_W = g\sqrt{z}\ \begin{array}{c}\includegraphics{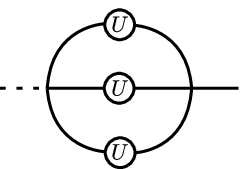}\end{array} + dgz\ \begin{array}{c}\includegraphics{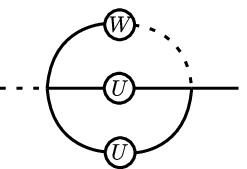}\end{array} = g\sqrt{z}\,U^d + dgz\,W\,U^{d-1}.\label{sigmaW}
\ee
As $\Sigma_V$ and $\Sigma_W$ are independent of $V$, and $\Sigma_W$ is only linear in $W$, one can extract from \eqref{getW} $W$ in term of $U$,
\be
W(U) = \frac{g\sqrt{z}\,U^{d+1}}{1-(d+1)gz\,U^d}.
\ee
This provides an expression $\Sigma_W(U)$ by inserting it into \eqref{sigmaW}. Then, one uses \eqref{getV} to extract $V$ in terms of $U$,
\be
V(U) = \frac{1+ W(U)\,\Sigma_W(U)}{1-gz\,U^d}.
\ee
This way, one gets rid of $V$ and $W$ easily and reduces the algebraic system to a single equation, i.e. \eqref{getU} where $W, V$ are functions of $U$. It is useful to change $U$ to $u = z g U^d$. Then the equation becomes linear in $g^{1/d}$,
\be \label{gdeu}
z^{\frac{d+1}{d}}\,g^{\frac{1}{d}} = u^{\frac1{d}}\ \frac{du^3 + [(d+1)^2 z+1] u^2 - [2(d+1)z+1] u +z}{(du+1)((d+1)u-1)^2}
\ee
The continuum limit is defined just like for pure random lattices as the points where $u$ becomes singular, i.e. $\partial g/\partial u=0$. This is equivalent to the stationarity of $g^{1/d}$ which reads
\be \label{zdeu}
\frac{\partial g^{\frac1{d}}}{\partial u}=0\quad\Leftrightarrow\quad
z = u\,\frac{(u-1)^2 + d^3 u^2 (u^2-2u+2) +d(u-1)(2u^2-1)+d^2u +d^2 u^2(u^2+u-3)}{\bigl((d+1)u-1\bigr)^3\,\bigl(d(d-1)u-1\bigr)}.
\ee
This gives the activity as a function of the parameter $u$. Plugging it back in \eqref{gdeu} we also get the coupling $g$ as a function of $u$. Therefore we obtain the line $(z_*(u),g_*(u))$ of the continuum limit as a parametrized curve.

\begin{figure}
 \subfigure[The solution of the model in the plane $(g,U)$ at $d=3$ for different values of $z$. The red line represents the continuum limit parametrized by $z$ and the black dot is at $z=0$.]{\includegraphics[scale=0.6]{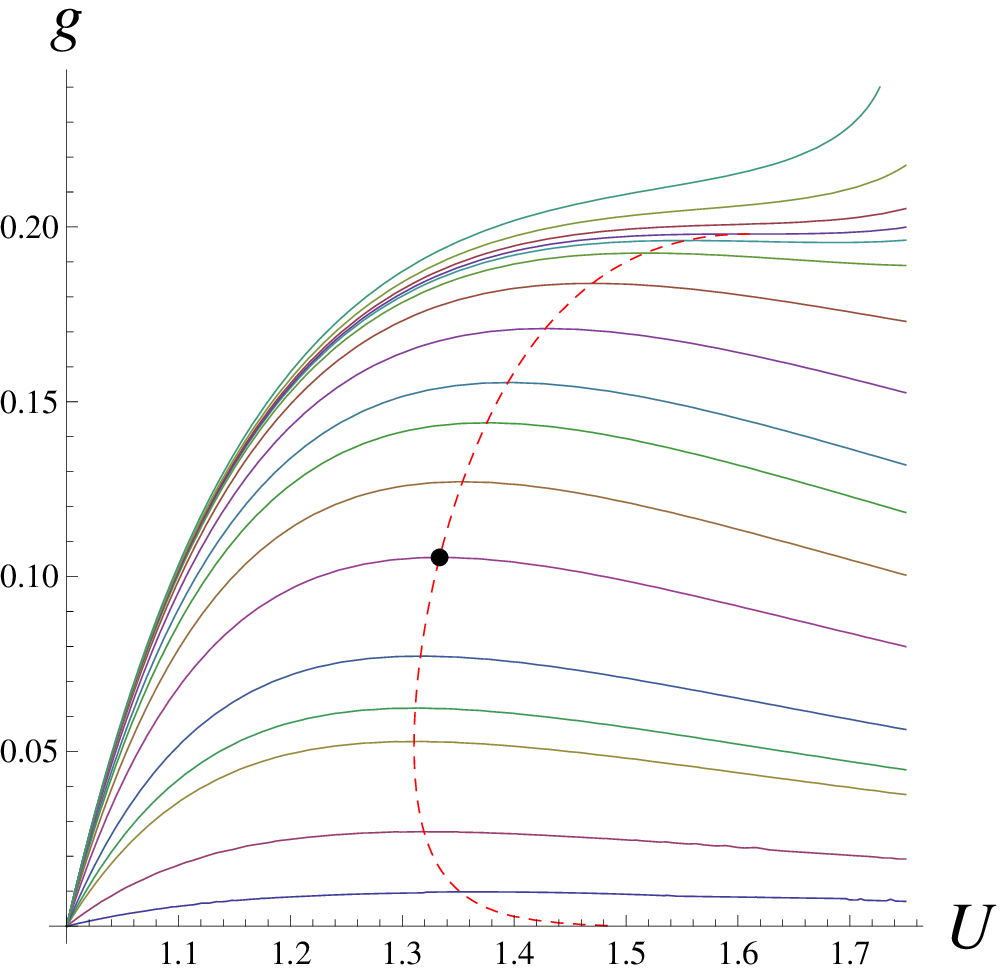} \label{fig:gdeU}} \hspace{2cm}
 \subfigure[The line of the continuum limit meets the set of local minima at the black dot for some critical activity $z_c$. ]{\includegraphics[scale=0.6]{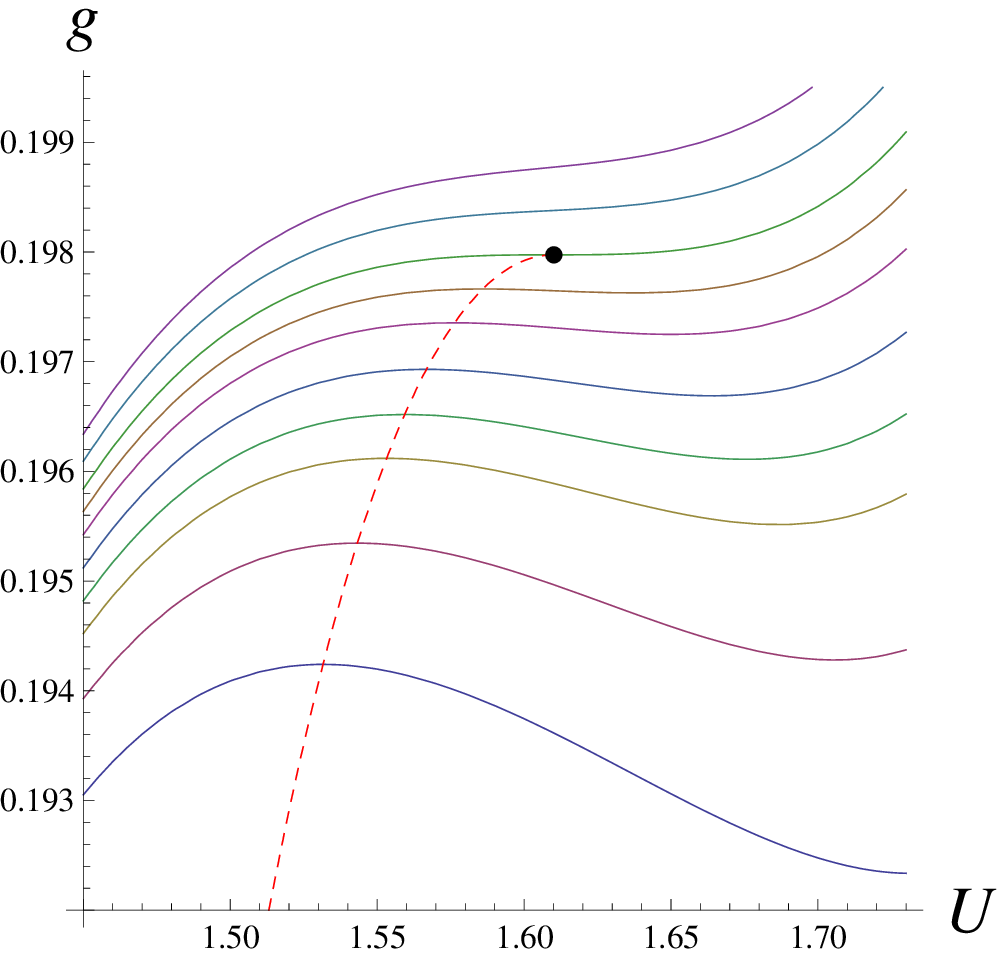} \label{fig:gdeUzoom}}
 \caption{ }
\end{figure}

To get a picture of this line, one draws \eqref{gdeu} in the plane $(g,U)$ for different values of the activity $z$, see Fig. \ref{fig:gdeU} at $d=3$ with $z$ between $-0.0935$ and $4$. Starting from $z=0$ one follows the local maxima obtained for different activities, represented as the red dashed line on Fig. \ref{fig:gdeU}. At fixed $z$, the first derivative with respect to $U$ vanishes and the second does not. Hence,
\be
\gamma = 1/2,
\ee
for small enough activities. This is the behavior of pure random lattices, as expected when dimers are not critical.

\subsection{Multicritical behavior}

Let us focus here on $d=3$. It can be observed on Fig. \ref{fig:gdeU} that for large enough activities there is no local maximum anymore, so that the line of the continuum limit stops somewhere. To understand this, one should notice that in addition to the set of local maxima which defines the physical continuum limit, there is a set of local, unphysical minima for each $z$, which actually come closer and closer to the physical line when $z$ is varied. This can be seen on Fig. \ref{fig:gdeUzoom}, where it becomes clear that there exists a critical value $z_c$ of the activity where the physical line meets the unphysical set of local minima. At this point not only the first derivative of $g$ with respect to $U$ vanishes, but also the second derivative. Hence performing an expansion around this point at $z=z_c$ fixed we find
\be
g - g_*(z_c) \sim A\ (U-U_c)^3 \qquad\Rightarrow\qquad U_{\text{sing}} \sim (g-g_*(z_c))^{1/3}\qquad \Rightarrow\qquad \gamma=2/3.
\ee
This means that the universality class of the continuum limit is not the one of pure lattices anymore. It is changed because dimers become critical at this activity $z_c$. Beyond $z_c$ the solutions to $\partial g/\partial U=0$ move to the complex plane.

Analytically this behavior take place when $\partial z_*/\partial u=0$. Taking the derivative of \eqref{zdeu} and after removing an obvious root at $u=-1/d$ we get
\begin{multline}
1+d +(-2+d+3d^2)\,u + (1-6d-6d^2+3d^3-4d^4)\,u^2 + (5+2d+2d^2+9d^3-2d^4)\,d\,u^3 \\+ (-1+d-3d^2-5d^3)\,d\,u^4 + (d-1)(d+1)^2 d^2\,u^5 = 0.
\end{multline}
This is a polynomial of degree 5 for which we have not found any obvious root. Hence, they have to be evaluated numerically. At $d=3$, the real roots are $-0.075, 0.185, 1.37$. The relevant one is the first, which corresponds to $z_c \approx -0.09$, in agreement with the plot \ref{fig:gdeUzoom}. One also gets $g_*(z_c)\approx 0.198, U_*(z_c)\approx 1.61$. Notice the critical activity is negative, as expected from the 2d case for non-unitary critical models. (For positive activities, one sees in Fig. \ref{fig:gdeU} that the solution of the model goes to $g=0$.)

Other interesting exponents are the ones which characterize the singularity when approaching it in the continuum limit. We move towards $z_c$ following the critical line $(z_*(u),g_*(u))$. Using some Taylor expansions it is found that $z_*$ and $g_*$ both start quadratically with $(u-u_c)$, therefore
\be
g_*(z) \approx g_*(z_c) + b\,(z-z_c)^{3/2} + \text{regular terms}.
\ee
To get the free energy in the continuum, we make use of the Schwinger-Dyson equations \eqref{SD} and find
\be
\frac{dF}{dz}_{|z_c} = \frac{\partial F}{\partial z} + \frac{dg_*}{dz}\,\frac{\partial F}{\partial g} \sim B\ \sqrt{z-z_c},
\ee
for some constant $B\approx -126.9$.

\section{Mapping to ``color-sensitive hard dimers'' on random branched polymers} \label{sec:mapmelons-trees}

It is known \cite{graph-combinatorics-difrancesco} that planar graphs can be mapped to branched polymers, i.e. trees. Similarly melonic graphs are in one-to-one correspondence with a family of branched polymers. This is actually one way to understand the series of multicritical exponents of random matrix and tensor models which go like $\gamma = \epsilon -1/m$ where $\epsilon=0,1$ and $m\geq2$.

\begin{figure}
 \includegraphics[scale=0.75]{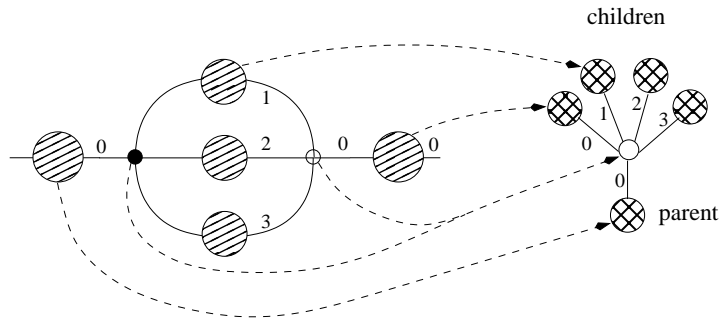}
 \caption{\label{fig:mapmelons-trees} The map from melonic graphs to branched polymers sends pairs of vertices to tree vertices. Every line of the graph is faithfully represented in the tree. The parent of a tree vertex comes from a melonic insertion connected to the black vertex, and children of the tree vertex represent melonic insertions connected to the white vertex.}
\end{figure}

While a tree is obtained from a planar graph by cutting lines, the mapping of melonic graphs is more abstract and goes as follows. First let us define a \emph{melon} as a \emph{1PI connected 2-point subgraph}. At any scale, a portion of a melonic graph looks like Fig. \ref{fig:mapmelons-trees}, i.e. a melon with left and right insertions on its external lines. One has two vertices connected by $d$ lines, with all colors but, say, $0$, and with (melonic) 2-point insertions on each line: this is a \emph{melon}. One also has some (melonic) 2-point insertions on the ``external'' lines to the melon, with color $0$. We map the two vertices to a single tree vertex and the lines of the graph to tree lines. The latter inherit the colors of the melonic lines. The tree vertex has a line of color $0$ which comes from its parent: it represents the line of color $0$ adjacent to the black vertex of the melon. Then the lines adjacent to the white vertex are mapped to lines connecting the tree vertex to its children.

The process can be repeated. Indeed every melonic 2-point insertion appearing in Fig. \ref{fig:mapmelons-trees} is itself of the type of Fig. \ref{fig:mapmelons-trees}. This way one associates to any 2-point melonic graph a branched polymer with vertex degree $(d+2)$, and technically known as a connected, \emph{rooted $(d+1)$-ary tree}.

\begin{figure}
 \subfigure[A dimer, represented as a thick blue line, on the line of color 1 (or 2 or 3) is mapped to a dimer on the tree line of the same color. In this case the hard-core repulsion is preserved.]{\includegraphics[scale=0.65]{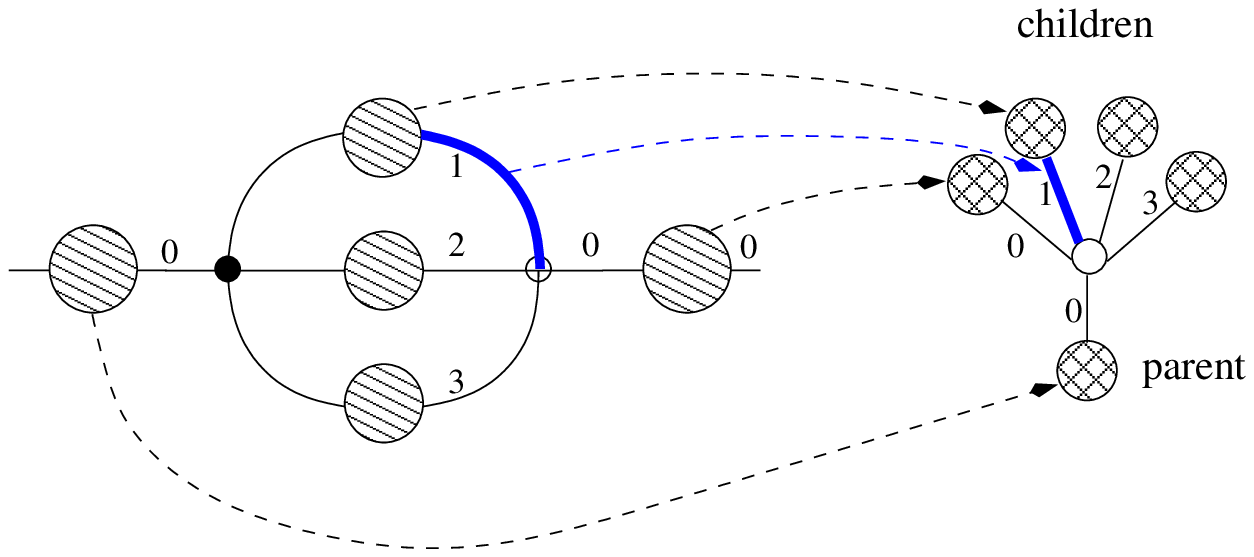} \label{fig:mapmelons-trees-dimers}}
 \subfigure[Two dimers on the external lines of a melon, here with color $0$, are mapped to lines connected to the same tree vertex. Therefore the dimers are hard-core except when they occupy lines of the same color.]{\includegraphics[scale=0.65]{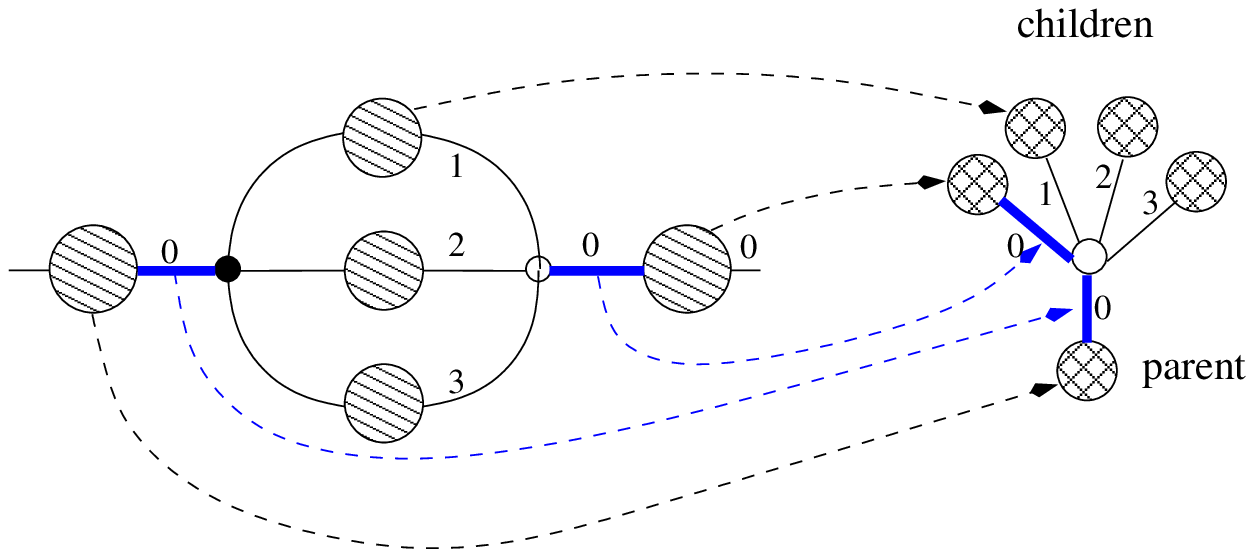} \label{fig:mapmelons-trees-dimers2}}
 \caption{Dimers on melonic graphs are mapped to ``color-sensitive hard-core dimers'' on branched polymers.}
\end{figure}

Now we can put dimers on melonic graphs and look at their mapping onto branched polymers. A dimer on the line of color 1 touching the white vertex on Fig. \ref{fig:mapmelons-trees-dimers} is mapped to a dimer on the line of color 1 on the tree. As no other dimer can touch the white vertex, no other dimer can touch the corresponding tree vertex. It goes the same way for dimers on the lines of colors 2 and 3, so it might seem that hard dimers are mapped to hard dimers. However, this is not exactly true as we need to look at the effect of dimers on the lines of color 0. This case is shown on Fig. \ref{fig:mapmelons-trees-dimers2}. As one dimer touches the white vertex and another one touches the black vertex, they are mapped to dimers on the two tree lines of color 0 which actually meet at the same tree vertex.

Therefore, the equivalent problem on branched polymer is a model of color-sensitive dimers. If a tree vertex has its parent of color $i$, then dimers on the children lines of color $j\neq i$ have their hard-core repulsion preserved and cannot touch. But dimers carried by lines of the same color are allowed to touch each other.

Then the 2-point function $U$ calculated in the section \ref{sec:critical-dimers} is the generating function for these color-sensitive dimers on random, connected, rooted $(d+1)$-ary trees. Branched polymers become critical along the red dashed line of Fig. \ref{fig:gdeU} which defines the thermodynamic limit in the grand-canonical ensemble of random branched polymers. A multicritical behavior is a point where trees and dimers become critical together, which happens at the critical activity $z_c$ as shown in Fig. \ref{fig:gdeUzoom}.

\section{More dimers} \label{sec:moredimers}

The parameter $z$ has allowed us to observe some multicritical behavior. With more parameters, one can actually achieve higher order multicritical points as shown in \cite{uncoloring}. An interpretation of such behaviors has been given in \cite{Bonzom:2012sz} in terms of generalized dimers coupled to random lattices. However the free energy there was a generating function for melonic graphs coupled to dimers at fixed values of $dV/2-|D|$, mixing the number of vertices and the number of dimers. This is not exactly what one would naturally consider: as the thermodynamic limit is defined by large values of the number of vertices $V$, one would like to work with a generating function which counts independently the number of vertices and the number of dimers. We show in this section how to write such a model, by a simple change of definition of the couplings in the model of \cite{Bonzom:2012sz}.

\begin{figure}
\subfigure[A graph dual to a triangulation at $d=3$ (since nodes are four-valent) with a set of two dimers on the left, a single dimer in the center and a set of three dimers on the right.]{\includegraphics[width=5cm, height=2cm]{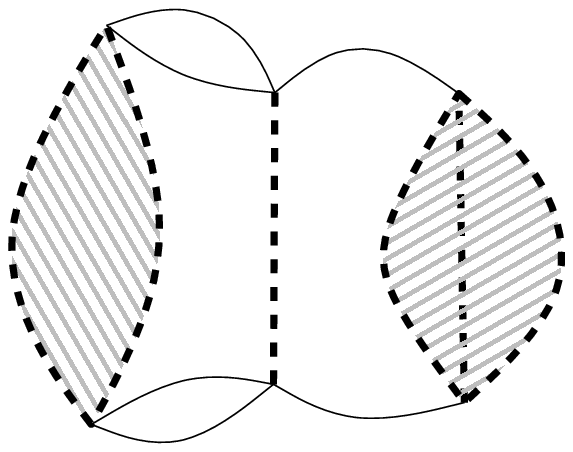} \label{fig:dimers}}\hspace{2cm}
\subfigure[ The dimer parts of the action can be drawn with internal (dashed) lines which represent dimers and internal faces (2 lines and 1 face for the set of two dimers on the left, 3 lines and 3 faces for the set of three dimers on the right). The effective interaction only sees the external legs carrying some $T$ and $\bar T$.]{\includegraphics[scale=0.65]{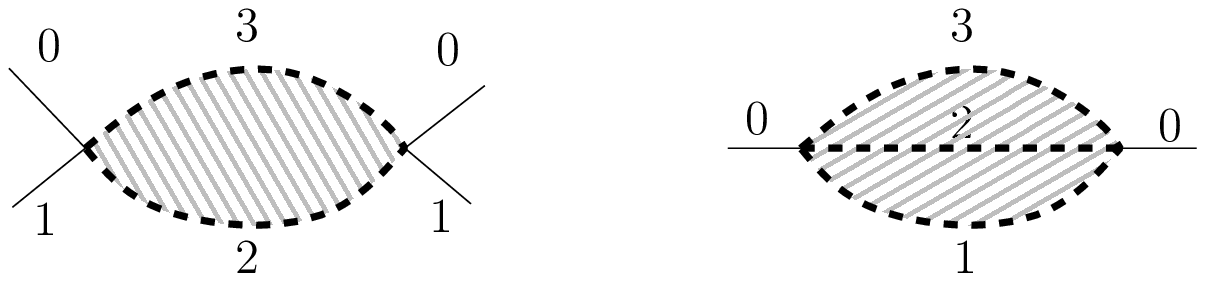} \label{fig:2-dimer}}
\caption{  }
\end{figure}

The main idea of the model is to allow several dimers to touch if they connect the same two vertices, as shown in Fig. \ref{fig:dimers}, \ref{fig:2-dimer}, and to assign to a set of $k$ dimers between two nodes some activity $z_k$.
These new terms are introduced in the action as effective vertices which depend on the external lines (and their colors) which join the two nodes. For two nodes connected by $d+1-k$ dimers, there are $k$ external lines corresponding to tensors $T^{(c_i)}$ on the black vertex, for some choice of colors $(c_i)$, and $k$ lines with the same colors for $\bar T^{(c_i)}$ on the white vertex. This gives
\be
S^{\{c_1,\dotsc,c_{k}\}}_{\rm dimers} = - g\,z_{d+1-k}\,N^{\frac{(1-k)(d-k)}{2}}\ \tr\, \prod_{i=1}^k T^{(c_i)} \bar T^{(c_i)}.
\ee
The index contractions are denoted with a trace, as it is otherwise a bit cumbersome. $S^{\{0,1\}}_{\rm dimers}$ and $S^{\{0\}}_{\rm dimers}$ are represented for $d=3$ on the figure \ref{fig:2-dimer}. The terms with $k=1$ can be seen as corrections to $S_{\text{quad}}$. The scaling with $N$ is chosen so that it is exact the same as gluing a node from $S_+$ and a node from $S_-$ via $d+1-k$ propagators, and hence $(d+1-k)(d-k)/2$ faces. The weight $g$ comes from one $\sqrt{g}$ for the white vertex and one for the black vertex.

The free energy of the model, at leading order, is a sum over melonic connected graphs
\be
\Xi_{\text{dimers}}(z,g) = \sum_{\{G_{2V}\}} g^{V} \sum_{D\in\mathcal{D}(G_{2V})} \prod_{k=2}^{d} z_{k}^{|D_k|},
\ee
where $|D_k|$ is the number of sets of $k$ dimers joining two nodes. In contrast with the model of \cite{Bonzom:2012sz}, $g$ now really controls the number of vertices of the lattice. For simplicity, we assume that $z_1=0$, i.e. we do not allow the standard hard-core dimers considered in the previous section.

Following the reasoning of \cite{Bonzom:2012sz}, similar to that of the above section, one has $U=1/(1-\Sigma)$ and $\Sigma$ can be obtained in the melonic sector as
\be
\Sigma = g\,U^d + g\sum_{k=0}^{d-2} \binom{d}{k} z_{d-k}\,U^k,
\ee
hence
\be
\frac1{g} = \frac{U^{d+1} + \sum_{k=0}^{d-2} \binom{d}{k} z_{d-k}\,U^{k+1}}{U-1}.
\ee
The continuum limit corresponds to the boundary of the analyticity domain of $U$ in $g$ where $\partial(1/g)/\partial U=0$. This reads
\be \label{z_d}
z_d = d\,U^{d+1} - (d+1)\,U^d +\frac{d(d-1)(d-2)}{2}\,z_2\,U^{d-1} + \sum_{k=1}^{d-2}\Bigl[(k-1)\binom{d}{k-1}\,z_{d+1-k} - (k+1)\binom{d}{k}\,z_{d-k}\Bigr]\ U^k.
\ee
This equation has the form $\Phi(U,z_k)=0$ and implies that for generic values of the couplings it can be locally solved to get $U$ in terms of the activities $z_2,\dotsc,z_d$. Then the entropy exponent is the one of pure random melonic lattices $\gamma=1/2$.

However, \eqref{z_d} is not invertible anymore if $\partial\Phi/\partial U=0$ which is here equivalent to $\partial z_d/\partial U =0$. When this is true together with \eqref{z_d}, the continuum limit becomes multicritical with $\gamma=2/3$ like in the previous section. Higher order multicritical points are also reachable when the first $m$ derivatives of $g$ vanish at the same time, $\partial^k(1/g)/\partial U^k =0$ for $k=1,\dotsc,m$. Thus our conlusion is similar to that of \cite{Bonzom:2012sz}: this simple model inspired by dimers exhibits multicritical behaviors.

In \cite{Bonzom:2012sz}, an example of a phase transition, i.e. a point where the physical solution to \eqref{z_d} meets another solution and that they exchange their roles, was presented for $d=6$. Here we want to pay attention to $d=3$ instead and compare with the hard dimers of the previous section. The equations for a multicritical point are
\be
\frac1{g} = \frac{U^4 +3 z_2\,U^2 +z_3\,U}{U-1},\qquad 
z_3 = 3\,U^4 - 4\,U^3 + 3z_2\,U^2 - 6z_2\,U,\qquad
z_2 = -2\,U^2,
\ee
where we have ignored the possibility that $U=1$. Clearly there can not be higher multicritical points as $\partial z_2/\partial U=0$ would imply $U=0$. Also notice that $z_2$ must be negative, in agreement with the expected non-unitarity of the dimer critical point.

\begin{figure}
 \includegraphics[scale=0.6]{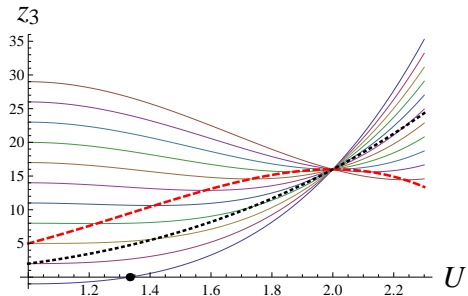}
 \caption{\label{fig:z3deU} The plot of $z_3(U)$ which characterizes the continuum limit, for different values of $z_2=-10,\dotsc,0$. The black dot is the continuum limit at $z_2=z_3=0$. The red dashed line is the multicritical line where $\partial z_3/\partial U=0$ and the black dotted line is the set of points where $1/g=0$.}
\end{figure}

Fig. \ref{fig:z3deU} plots $z_3$ as a function of $U$ for different values of $z_2=-10,\dotsc,0$. The curves span a two-dimensional region of the plane. The black dot is the continuum limit at $z_2=z_3=0$, where $U=4/3$. The red dashed line represents the set of multicritical points where $\partial z_3/\partial U=0$, so in principle it can be reached smoothly from the initial configuration $z_2=z_3=0$.

However, there is another interesting curve which we have plotted as a black dotted curve and corresponds to points where the coupling $1/g$ vanishes, $z_2=-U^2$. It is not clear to us what the physical meaning of these points is, but it is clear that the multicritical line can not be reached from $z_2=z_3=0$ without crossing such a point where $1/g=0$.

\section{Conclusion}

We have shown that hard dimers on dynamical triangulations, generated by random tensors in the large $N$ limit, are solvable and exhibit a critical point in the thermodynamic limit. This is quite similar to the Yang-Lee singularity observed in the two-dimensional case, and in particular it also happens for a negative activity. The missing element of our analysis is obviously the identification of the universality class in term of a field theory. But we do not know yet the degrees of freedom in the continuum which emerge from the sum over melonic graphs. This is certainly an important challenge in the future. The present paper is a contribution to gain insights from the phenomenology of lattice models coupled to these melonic graphs.

If one is able in the future to give a field theory description of the critical point, one could then consider a sort of KPZ correspondence to relate critical phenomena on flat space to critical phenomena coupled to geometric fluctuations. It is not clear how this correspondence could work in dimension higher than two. For example, it was shown in \cite{ising-colored} that the Ising model has no phase transition on dynamical melonic graphs while we do know it has one on flat space for any $d$. From this perspective a KPZ correspondence should be quite violent. However the present paper shows that non-unitary critical points can be observed, which makes the KPZ correspondence hypothesis more reasonable. Further work is needed to decide whether (non-)unitarity plays any role.

Another challenge is to use tensor models beyond the melonic leading order. In matrix models, it is possible to do so using the Schwinger-Dyson equations for instance, to solve the higher genus contributions from the planar ones. For tensors, the algebra generated by the Schwinger-Dyson equations has been presented in \cite{Gurau:2012ix} and contains an interesting sub-algebra generated by melonic observables, equivalently $d$-ary trees, that we propose to call the Gurau algebra \cite{Gurau:2011tj}. It turns out that \emph{at leading order} this sub-algebra reduces to (the positive part of) the Virasoro algebra as shown in \cite{uncoloring}. However this is not true at higher orders in the $1/N$ expansion. The consequences of these different symmetries on critical behaviors should be studied too.

\section*{Acknowledgements}

V.B. would like to thank Jan Ambj\o{}rn for discussions and for his enthusiasm for talking about statistical mechanics on dynamical geometries. Both authors are thankful to John Berlinsky who made H.E.'s stay in Perimeter Institute and hence the completion of this work possible.

V.B. is also thankful to Fabien Alet, Sylvain Capponi, Matthieu Mambrini and Pierre Pujol for their rich explanations about dimers in condensed matter physics.

Research at Perimeter Institute is supported by the Government of Canada through Industry Canada and by the Province of Ontario through the Ministry of Research and Innovation.


\end{document}